\documentclass[twocolumn,showpacs,preprintnumbers,amsmath,amssymb]{revtex4}

\usepackage{graphicx}
\usepackage{dcolumn}
\usepackage{bm}
\usepackage{amsmath}
\usepackage{amssymb}

\begin{document}

\preprint{}

\title{Temperature dependence of critical 
       velocities
       and long time decay of supercurrents}
 
\author{Yongle Yu}
\affiliation{State Key Laboratory of Magnetic Resonance
  and Atomic and Molecular Physics, Wuhan Institute of Physics and Mathematics,
 CAS,  Wuhan 430071, P. R. China}


\date{April, 2008}
\begin{abstract}

We show that a  microscopic theory of
superfluidity, based on the properties of 
the many-body spectrum of a superfluid, 
can explain naturally  the temperature 
dependence of  critical velocities
and the long time decay of supercurrents.
 

\end{abstract}
\maketitle

Landau gave the first understanding of 
persistent currents in $^4$He 
 by  
 relating this  phenomenon
with 
properties of the quasiparticle spectrum
 of the system \cite{landau}. 
Bogoliubov explained the linear dispersion of
the quasiparticle spectrum at small momenta 
with the assumption of BEC fraction in a superfluid 
\cite{bogoliubov},
 and thus
supported Landau's theory. This 
microscopic picture of superfluid 
is, however, not invoked 
 in understanding some later important  
 observations of superfluids, such as,  ${\star}A$) 
 the temperature
 dependence of critical velocities \cite{reppyal, hess2},
   and ${\star}B$) 
 the long time decay
 of supercurrents \cite{reppydecay}.  
 It is natural to wish that a theory of
 superfluidity can account for these
 observations besides explaining superfluidity, for
 the purpose of a unified picture to 
 describe superfluidity
 and its related properties.
 In this paper, we show that
a rather simple microscopic
theory of superfluidity  
can accomplish such a wish. 
We also make comparisons between this
 theory and Iordanskii-Langer-Fisher
(ILF) theory \cite{ILF}, which is formulated to understand
the temperature dependence of critical velocities
and in which multiple assumptions of 
vortex rings are essential.

\begin{figure}
\begin{center}\includegraphics
{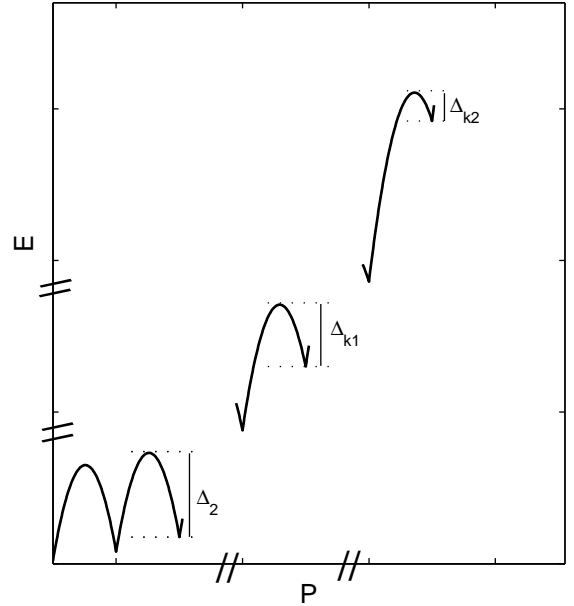}\end{center} 
\caption{ A many-body dispersion spectrum of 
a $1D$ Bose system. The barrier height decreases
with increasing of the momentum
at the corresponding minimum. The height of the barrier
for the $k${\it{-th}}
 minimum next to the ground state regime 
  is $\Delta_k= \Delta_0
 (1- k/(v_c MR/\hbar))^2 $ with
the momentum at the minimum is $ P= k N \hbar/R   $. 
The minima disappear beyond a certain momentum.}
\label{dispersion}
\end{figure}

 
This microscopic theory relates superfluidity
with the properties of the 
 many-body spectrum of a superfluid.
The main conclusions of the theory 
are the following: {\it i)} The 
many-body dispersion spectrum of
a superfluid $E= E(P)$, where $E$ is the lowest
 many-body eigen-energy at given momentum $P$, 
 is not a monotonic
function of $P$; there exist energy
barriers in the many-body dispersion spectrum which
separate and prevent some current-carrying states from
decaying \cite{bloch, leggett, yu}
 (see Fig. \ref{dispersion}).  {\it ii)}
 the existence 
of the energy barriers is due to Bose 
exchange symmetry \cite{yu}.  {\it iii)} The height 
of barriers will decrease with increasing momentum
(velocity) of the corresponding supercurrents;
and beyond a certain
velocity, the barriers disappear and the system 
dissipates
its momentum like a normal system  \cite{bloch, yu}. 
We shall show that with
this knowledge of the many-body spectrum, 
  the observations (${\star}A$, ${\star}B$)
can be naturally explained.
We  first use the spectrum of
a one-dimensional 
superfluid to  
illustrate these observations,
and later discuss 
higher dimensional superfluids of which
the gross structures of the 
spectra are also the same as what specified above.

We consider  $N$  Bose particles 
moving in a ring with a radial size of
$R$.
The Hamiltonian has the form of
\begin{equation}
    H= - \sum_{i=1}^{N} \frac{\hbar^2}{2MR^2}
    \frac{\partial^2} {\partial 
  \theta_i^2}+  \sum_{i<j}^N V(|\theta_i-\theta_j|),
\end{equation}

where $\theta_i$ is the
angular coordinate of the $i$th particle, $M$ is
 the mass of a particle 
and $V$ is the repulsive interparticle interaction. 
$V$ can be either short-range or zero-range. 
The many-body spectrum can be classified by their
angular momenta $L$ (in following we replace $L$
by a corresponding momentum $P= L/R$, considering
the equivalent system moving in a straight line with
periodic boundary condition). Due to Galileo 
invariance, the full spectrum 
can be obtained if
the part at momentum regime $0 \leq P \leq   N \hbar/R$
 is  known  \cite{bloch, yu}. Specifically, if one denotes
 the $n${\it{-th}} level at momentum $P$ by $e_n(P)$, then
\begin{equation}
 e_n(P +  kN \hbar/R)= e_n(P) + ((P +   k N \hbar/R)^2 - P^2)/2NM,
\end{equation}
where $k$ is an integer.

The many-body
 dispersion spectrum, as a function of $P$
at regime $0 \leq P \leq   N \hbar/R$, relative to
the ground state energy,  can be written in
 a form of
 \begin{equation}
  E(P) \equiv E_{ex}(P) +  \hbar P/2MR ,
  \end{equation}
  where  $ \hbar P/2MR $ is  a (trivial) part of kinetic 
  energy,  and where 
 $E_{ex}(P)$ 
is symmetric (in $E-P$ plane) to
 the line $P= N \hbar  /2R$ and reaches its
maximum at $P= N \hbar /2R$. Numerical results \cite{yu, lieb}
 suggest that
$E_{ex}(P)$ is parabolic, i.e.,
$E_{ex}(P)= v_c  (  N \hbar /R -P  ) P / (N \hbar/R) $,
we shall use this  form of  $E_{ex}(P)$
for discussions. 

With the knowledge of $E_{ex}(P)$, one 
can obtain the full 
many-body dispersion spectrum and
realize the following feature
of this spectrum.  i) for a positive integer
$k < v_c MR/ \hbar $, there is 
an energy barrier in the dispersion spectrum
within the momentum regime
  $(k-1)N\hbar/R \leq P \leq kN \hbar/R$.
ii) the height of this barrier is 
$\Delta_k= \Delta_0 (1- k/(v_c MR/\hbar))^2$,
where $\Delta_0= v_c N \hbar/4R $ 
(see Fig. \ref{dispersion}). 

\begin{figure}   
\begin{center}\includegraphics
{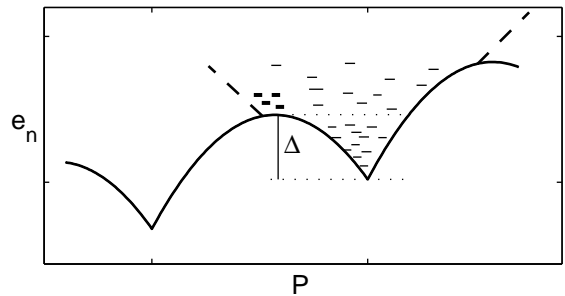}\end{center} 
\caption{With a sufficient small temperature, a system 
can stay in a valley for a long time. The many-body levels
(bars) in the valley region, which is bordered by the valley and 
roughly bordered by 
the thick dashed line, will reach a local
quasi thermal equilibrium
and the probability at a level with energy $e_n$ is determined
by the usual law,  $\rho \propto e^{-e_n/k_B T} $. With a finite 
$T$, the states  (thick bars) near the 
next lower valley are partially accessed 
and subsequently decay to the next valley, thus causing a 
leak of  the system from this valley to the lower one.}
\label{valley}   
\end{figure}

Each valley of the dispersion spectrum 
can 'trap' metastable states
of the system given that the temperature is lower enough. 
for a certain valley, we  consider the quasi thermal
equilibrium distribution of many-body
levels in the region plotted in Fig. \ref{valley}, then
the possibility for the system staying at 
the levels lower than the energy barrier $\Delta$  is
that
\begin{equation}
   \gamma = \frac {\sum_i e^{- (e_i- e_o) /k_B T}} 
          {\sum_j e^{- (e_j - e_o)/ k_B T}} \;\;\; (e_i - e_o < \Delta )
\end{equation}
Where $i,j$ refers to levels in the
 region, $k_B$ is the Boltzmann constant,
  and $e_o $ is the lowest eigen-energy
 at the valley, i.e., the local minimum of the dispersion
 spectrum.  

 At any small but finite $T$, $\gamma$ is smaller than unity.
 It is with some possibilities that
  some states, which are near the edge of the valley regime and close to
 the next lower valley (thick bars in Fig. \ref{valley}),  can be reached,
 and subsequently
 these states can decay with ease to the states in the next 
 lower valley, (i.e., they are kind of 'doorway' states). This process 
 is then repeated and
 eventually the system transfers from this
 valley to the next valley. When  $\gamma $ is the close
 to one, this transferring process is slow, which, nevertheless,
 leads to the decay of the supercurrent at the large time
 scale  \cite{reppydecay,2dfilm}. 
 
 When $\gamma $ exceed a certain value $\gamma_c$ close to
 unity, one can  ignore the long time slow decay of 
 supercurrents if only relatively short time scale 
 is concerned.
  the value of $\gamma $
 depends on the many-body spectra, $T$ and  $\Delta_k $. 
 If
 one roughly estimates that  $\gamma = \gamma_c $ at $T = c \Delta_k$
 at the $k${\it{-th}} valley,
 where $c$ is a constant,
 one then finds  the highest possible  velocity
 of supercurrents at $T$  is given by 
 $v_c(T)= v_c (1- \sqrt{T/T_\lambda}) $ where
 $T_\lambda$ is the transition temperature 
 \cite{specifications}.
 One shall note
 that the local spectra at different valleys are not 
  the same, for example, the shapes of the 
   valleys become more asymmetric with valleys
    further away from the ground state regime. 
     For this reason, the
  rough temperature-critical-velocity relation 
  could be rendered to take a form of
  \begin {equation}
   v_c(T)= v_c (1- \sqrt{\alpha(T)T/\alpha(T_\lambda)T_\lambda}) 
   \end{equation}
   where $\alpha(T)$ is a slowly varying function of $T$.
 
  We thus illustrated the theoretical pictures of
  two observations (${\star}A$, ${\star}B$) using the many-body spectrum.
  One could note that the pictures are direct and unavoidable. 
  A spectrum  of a system is a fundamental
   property of the system, and one can map 
   physical processes to the transferring processes
  among the many-body levels of the system.

  Previously, ILF theory is constructed to explain the
  temperature dependence of critical velocities. It is 
  interesting to compare our theory with ILF theory. One
  can  see the following
  agreements between them: {\it a)} that
  there are energy barriers to prevent the decay of 
  currents at short time scale; {\it b)} the energy barriers
  can be 'overcame' by thermal excitations, which leads
  to long time decay of supercurrents; {\it c)} the energy
  barrier is a decreasing function of the velocity of
  the supercurrent. 
  
  The differences between two theories are the following:
  {\it a)} ILF involves multiple assumptions of vortex rings,
     such as their sizes, their dynamics, the creation
     and annihilation of vortex rings.  Our theory suggests
     that these assumptions  are not necessary \cite{doorway-state}.
   {\it b)} ILF theory determines the heights of the energy barriers, 
    using energetics of vortices, and suggests the heights  decrease
     linearly as the function of the velocity of the supercurrent,
   i.e., $ \Delta(v)= \Delta_0 - p_o v$ \cite{ILFformula}.
   these 'conclusions'  lead to 
    inadequacies of the theory in its quantitative
    description of some systems \cite{2dfilm}.
    They also
   lead to  inadequacies in quantitative description 
   at the temperature regime
   far below the transition point \cite{hess2}. Within
   our theory, the height of energy barriers are
   naturally determined by the many-body spectra for 
   which all low-lying eigenstates including the
    many-body dispersion states are relevant. 
   

  We shall discuss higher dimensional cases. We consider
  for example a system of $N$ particles
  in a tube with periodic boundary condition.  
  The section area  $\sigma$ of a tube could be
   much smaller than $R^2$, where $R$ is 
  the length of tube divided by $2 \pi$, but is 
  order(s) of magnitude larger
  than $a^2$,  where $a$ is the average interparticle
  distance. Again, with the knowledge of
   dispersion spectra in the momentum 
  regime  $ 0 \leq P \leq N \hbar /R$ \cite{bloch, yu, yu2},
   one can derive the 
  full dispersion
  spectra, particularly all local minima and 
  the energy barrier
  height associated with each minimum, due to the Galileo
  invariance. 
  
   \begin{figure}
\begin{center}\includegraphics
{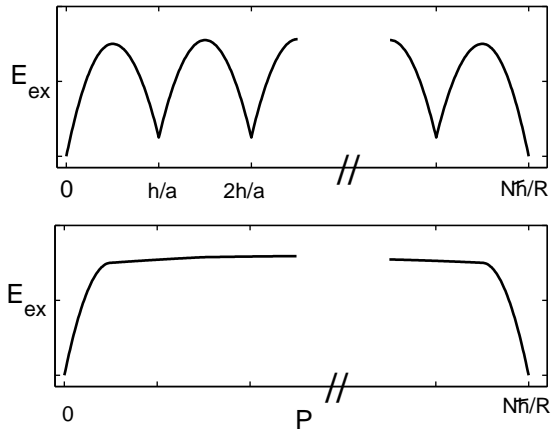}\end{center} 
\caption{ Possible scenarios of dispersion 
structure at the momentum regime $0 \leq P \leq N\hbar/R $. 
$ E_{ex}(P) \equiv E(P) - \hbar P /2MR$, where $E(P)$
is the dispersion spectrum. (Upper panel)
If certain conditions on boundary and on $N$ value 
are satisfied, there could be 
local minima at $h/a$, $2h/a$,... ($h= 2\pi \hbar$). 
(Lower panel)
In general cases, only the primary type of minima 
plays an important role in determining superfluidic properties.}
 
\label{highDE}
\end{figure}

  Beside the primary type of minimum at
  $P= N \hbar/R$, with the corresponding many-body state
  in principle allows a fraction of BEC, in \cite {yu2}
  we also find some local minima at $P= h/a, 2h/a, ...$,
  the existence of these local minima has sensitive
   dependences on boundary conditions and on the value of
   $N$ (see Fig. \ref{highDE}). It also 
   requires strong interaction.
   Once these conditions are not satisfied, 
   that type of
   minima  at $P= h/a, 2h/a, ...$ may 
    not exist or be very shallow, and only
    the primary type of minimum plays an essential role
    in superfluidity.
    
     
    In conclusion, we illustrate that a microscopic theory
    of superfluidity, based on the properties of a superfluid's
    many-body spectrum, explains naturally
    the temperature dependence of critical 
    velocities and the long time decay of supercurrents.
    We emphasis that these pictures of a superfluid
     are direct and unavoidable,  due to that
     physical processes can be viewed as changes
    of occupation probability of each  eigen-level 
    and the transfers among them.



\begin{thebibliography}{1}
\bibitem{doorway-state} Note that the doorway states approach
 the supercurrent states with increasing momenta
 in the plot of levels in the E-P plane. This suggests that 
 internal structures
 of the doorway states change along the dispersion curve,
  and can not be attached to a vortex ring structure 
  in the whole superfluidic 
 low momentum regime.
\bibitem{ILFformula} precisely, $\Delta(v) = E (r)- p_o(r)v $,
where $r$ is the size of a vortex and $E(r)$ and $p_o(r)$ depends
on $r$.
\end{thebibliography}
 \end{document}